\documentclass[a4paper, oneside, twocolumn, notitlepage, 10pt]{extarticle_ecoc}
\usepackage{ecoc}

\usepackage{orcidlink}
\usepackage{siunitx}
\DeclareSIUnit\bps{\bit\per\s}
\DeclareSIUnit\kbps{\kilo\bit\per\s}
\DeclareSIUnit\evt{evt\per\s}
\DeclareSIUnit\kevt{\kilo\evt}
\usepackage{pgfplots}
\pgfplotsset{compat=1.18}
\usepackage{xcolor}
\usepackage{hyperref}
\usepackage{comment}
\usepackage{tikz}
\usetikzlibrary{shapes.geometric, positioning, arrows.meta, fit}
\usepackage{anyfontsize}

\usepackage{eso-pic}
\usepackage{tcolorbox}
\AddToShipoutPicture*{%
  \AtPageUpperLeft{%
    \raisebox{-1.5cm}{ 
      \makebox[\paperwidth]{%
        \centering
        \begin{minipage}{1\textwidth} 
          \centering
          \scriptsize
          This paper has been accepted for the 2026 European Conference on Optical Communications (ECOC), Malaga, Spain. © 2026 OPTICA. \\
          Please cite it as:
          \textit{F. Nardo, P. Krötz, R. Barrios, and S. Randel, "Real-Time Event-Based Optical Camera Communication at 80 kbit/s over a 320 m Outdoor Link with 120° Field-of-View" in 2026 European Conference on Optical Communications (ECOC), Technical Digest Series (Optica Publishing Group, 2026), paper We4\_P107; }
          \url{https://doi.org/10.1364/ECEOC.2026.We4_P107}
        \end{minipage}
      }
    }
  }
}
\AddToShipoutPicture*{ 
  \AtPageLowerLeft{%
    \raisebox{0.3cm}{ 
      \makebox[\paperwidth]{%
        \begin{tcolorbox}[colframe=black, colback=white, sharp corners, width=\textwidth, ]
          \scriptsize
          © 2026 Optica Publishing Group and ECOCMALAGA2026 S.L. 
          One print or electronic copy may be made for personal use only. Systematic reproduction and distribution, duplication of any material in this paper for a fee or for commercial purposes, or modifications of the content of this paper are prohibited.
        \end{tcolorbox}
      }
    }
  }
}

\begin{document}
\selectlanguage{english}    


\title{
Real-Time Event-Based Optical Camera Communication \\ at 80 kbit/s over a 320 m Outdoor Link with 120\textdegree \ Field-of-View \\}%


\author{
    Francesco Nardo \orcidlink{0009-0006-5750-5837} \textsuperscript{(1)(2)}, 
    Peter Krötz \textsuperscript{(1)},
    Ricardo Barrios \orcidlink{0000-0002-2545-5568} \textsuperscript{(1)},
    Sebastian Randel\textsuperscript{(2)} 
}

\maketitle                  


\begin{strip}
    \begin{author_descr}

        \textsuperscript{(1)} Airbus Defence and Space GmbH, Munich, Germany,
        \textcolor{blue}{\uline{francesco.nardo@airbus.com}}

        \textsuperscript{(2)} Institute of Photonics and Quantum Electronics, Karlsruhe Institute of Technology, Karlsruhe, Germany.


    \end{author_descr}
\end{strip}

\renewcommand\footnotemark{}
\renewcommand\footnoterule{}


\begin{strip}
    \begin{ecoc_abstract}
       We demonstrate an event-based optical camera communication system using a \qty{120}{\degree} field-of-view sensor. In daylight, three transmitting nodes achieve an aggregate \qty{240}{\kbps}, with individual links reaching \qty{106}{\kbps} at \qty{1}{\m} and \qty{80}{\kbps} at \qty{320}{\m}.
       Our results, establish a new baseline for event-OCC.
    \end{ecoc_abstract}
\end{strip}


\section{Introduction}
Event-based Optical Camera Communication (event-OCC) utilizes Neuromorphic Vision Sensors (NVS), also known as Dynamic Vision Sensors (DVS) \cite{cazzatoApplicationDrivenSurveyEventBased2024}, for data transmission \cite{perez-ramirezOpticalWirelessCamera2019}. Compared to traditional Free Space Optical Communication (FSOC) receivers, event-OCC provides a wider Field-of-View (FoV) but lower data rates and sensitivity \cite{chowdhuryComparativeSurveyOptical2018}. 
Although kilobit-per-second (\qty{}{\kilo\bit\per\s}) rates are modest by FSOC standards, they are sufficient for critical telemetry and real-time situational awareness. 
For context, a \textit{240p} \qty{30}{fps} grayscale video stream compressed with the H.264/AVC standard typically requires bit-rates between \qty{300}{} and \qty{700}{\kbps}. The more efficient H.265/HEVC standard achieves comparable visual quality at bit-rates as low as \qty{150}{} to \qty{350}{\kbps} \cite{sullivanOverviewHighEfficiency2012a}.

While frame-based cameras are the standard OCC receiver, their bandwidth for Intensity Modulation/Direct Detection (IM/DD) techniques is typically restricted to less than $\qty{1}{\kilo\hertz}$ by limited frame rates. Strategies like Quaternary Amplitude Modulation (4-PAM) \cite{nguyenEnhancementDataRate2018} or spatial encoding (e.g., QR codes \cite{saeedOpticalCameraCommunications2019}) attempt to overcome this limitation.
In contrast to frame-based cameras, Neuromorphic cameras operate on differential signals, asynchronously recording intensity changes as event tuples $(x, y, t, p)$, where $x, y$ are pixel coordinates, $t$ is the microsecond-resolution timestamp, and $p$ is the polarity ($1$ for rising, $0$ for falling edges). This architecture is ideal for binary IM/DD schemes like On-Off Keying (OOK).
To the best of our knowledge, this work establishes a new performance baseline for event-OCC by achieving an unmatched combination of range and throughput. We report a data rate of \qty{80}{\kilo\bit\per\second} at a distance of \qty{320}{\m}. This result exceeds the highest reported data rates for long-range OCC. Furthermore, in a laboratory setting, we pushed the peak throughput to \qty{106}{\kilo\bit\per\second} at a short range. In multi-node operation, using three parallel incoming beams the system achieved an aggregate data rate of \qty{240}{\kbps}. In all cases, a Bit Error Ratio $BER \leq 10^{-3}$ was maintained.

\section{State of Art}

\begin{figure}[t]
    \centering
    \begin{tikzpicture}
\begin{axis}[
    width=8cm,
    height=5cm,
    tick label style={font=\tiny}, 
    label style={font=\tiny},
    xlabel={Distance [\qty{}{\m}]},
    ylabel={Data rate [\qty{}{\bit\per\s}]},
    xmode=log,
    ymode=log,
    xmin=0.1, xmax=1000,
    ymin=100, ymax=1000000,
    xtick={0.1, 1, 10, 100, 1000},
    xticklabels={0.1, 1, 10, 100, 1k},
    ytick={100, 1000, 10000, 100000, 1000000},
    yticklabels={100, 1k, 10k, 100k, 1M},
    grid=both,
    grid style={line width=.1pt, draw=gray!10},
    major grid style={line width=.2pt, draw=gray!30},
    legend pos=south west,
    legend cell align={left},
    legend style={nodes={scale=0.5, transform shape}}
]

\addlegendimage{only marks, mark=square}
\addlegendentry{Indoor}
\addlegendimage{only marks, mark=o}
\addlegendentry{Outdoor}
\addlegendimage{only marks, mark=triangle}
\addlegendentry{Multi-node}

\addplot[only marks, mark=square] coordinates {(5, 500)};
\node[below, scale=1] at (axis cs:5, 500) {\tiny Perez et al. 2017 \cite{perez-ramirezOpticalWirelessCamera2019}};

\addplot[only marks, mark=square] coordinates {(10, 16000)};
\node[below, scale=1] at (axis cs:10, 16000) {\tiny Shen et al. 2018 \cite{shenVehicularVisibleLight2018}};


\addplot[only marks, mark=square] coordinates {(0.5, 4000)};
\node[below, scale=1] at (axis cs:0.5, 12000) {\tiny Wang et al. 2022 \cite{wangSmartVisualBeacons2022}};
\addplot[only marks, mark=o] coordinates {(100, 500)};
\node[below, scale=1] at (axis cs:100, 500) {\tiny Wang et al. 2022 \cite{wangSmartVisualBeacons2022}};

\addplot[only marks, mark=square] coordinates {(0.5, 2500)};
\node[below, scale=1] at (axis cs:0.5, 2500) {\tiny von Arnim et al. 2024 \cite{vonarnimDynamicEventbasedOptical2024}};

\addplot[only marks, mark=square] coordinates {(1.5, 114000)};
\node[below, scale=1] at (axis cs:1.5, 114000) {\tiny Su et al. 2025 \cite{suMotionAwareOpticalCamera2025}};

\addplot[only marks, mark=o, transform shape=false] coordinates {(200, 60000)};
\node[below, scale=1] at (axis cs:200, 60000) {\tiny Sumino et al. 2025 \cite{suminoExperimentalDemonstrationEventbased2025}};
\addplot[only marks, mark=o, transform shape=false] coordinates {(400, 30000)};
\node[below, scale=1] at (axis cs:400, 30000) {\tiny Sumino et al. 2025 \cite{suminoExperimentalDemonstrationEventbased2025}};

\addplot[only marks, mark=triangle*, color=blue] coordinates {
    (1, 240000)
};    
\addplot[only marks, mark=square*, color=blue] coordinates {
    (1, 106000)
};
\addplot[only marks, mark=*, color=blue] coordinates {
    (320, 80000)
};

\node[color=blue, right, scale=1, font=\bfseries] at (axis cs:0.12, 170000) {\tiny This work $\longrightarrow$};
\node[color=blue, right, scale=1, font=\bfseries] at (axis cs:35, 90000) {\tiny This work $\longrightarrow$};

\end{axis}
\end{tikzpicture}
    \caption{Chronological progression of data rates versus distances in event-OCC research.}
    \label{fig:history}
\end{figure}
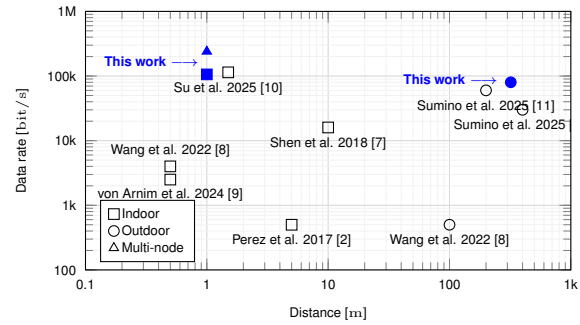

Neuromorphic cameras have been commercially available since 2001, following initial surveys on sensor development in the mid-1990s \cite{cazzatoApplicationDrivenSurveyEventBased2024,gallegoEventBasedVisionSurvey2022,tofighiSurveyEventbasedOptical2025}. However, their application in OCC has rapidly developed only within the last five years, driven by the increased availability of COTS embedded systems.
Prior research in event-OCC has largely focused on static links using Visible Light Communication (VLC) with LED transmitters. 

In 2017, Perez et al. demonstrated a \qty{500}{\bit\per\s} link over \qty{5}{\m} with $BER$ $< 10^{-4}$ \cite{perez-ramirezOpticalWirelessCamera2019}. 
Shen et al. reported a \qty{16}{\kbps} data rate at \qty{10}{\m} in 2018, though $BER$ performance was not evaluated \cite{shenVehicularVisibleLight2018}. 
By 2022, Wang et al. extended the transmission distance to \qty{100}{\m} at \qty{500}{\bit\per\s} with reported zero bit errors \cite{wangSmartVisualBeacons2022}. 
More recently, in 2024, von Arnim et al. introduced a frameworks for dynamic event-OCC environments at \qty{2.5}{\kbps} with  \qty{94}{\%} message accuracy at \qty{0.5}{\m} distance \cite{vonarnimDynamicEventbasedOptical2024}.
During the same year, Nakagawa et al. demonstrated a link over \qty{5}{\m} with $BER < 10^{-2}$ that included tracking capabilities for targets moving up to \qty{100}{\centi\meter\per\second} \cite{nakagawaLinkingVisionMultiAgent2024a}. Other research has explored bio-inspired communication protocols for drones, such as those modeled on the movement of bees \cite{nengboMoComMotionbasedInterMAV2025}.


Spatial encoding has further increased throughput, notably achieving the current event-OCC short-range (\qty{1.25}{\m}) baseline standing at \qty{114}{\kilo\bit\per\second} via dynamic QR codes, Su et al. 2025 \cite{suMotionAwareOpticalCamera2025}. 
Contemporaneous with this work, in October 2025, Sumino et al. reported a \qty{60}{\kbps} link at \qty{200}{\m} and \qty{30}{\kbps} at \qty{400}{\m} with $BER < 10^{-3}$ \cite{suminoExperimentalDemonstrationEventbased2025}. In March 2026, Dhillon et al. demonstrated simultaneous communication and tracking of a modulated LED at short range, achieving \qty{10}{\kbps} with a message accuracy of \qty{95}{\%} \cite{dhillonRealTimeOpticalCommunication2026}. The progression of experimentally demonstrated Event-OCC links is summarized in Fig. \ref{fig:history} together with the results demonstrated in this work.

Finally, modulation remains a critical challenge because neuromorphic sensors require temporal intensity changes to generate events. To ensure regular edge transitions for clock recovery, previous works utilized 8b/10b encoding \cite{widmerDCBalancedPartitionedBlock8B1OB, suminoExperimentalDemonstrationEventbased2025} or specialized schemes like event interval modulation \cite{suminoEventIntervalModulation2025}, n-pulse modulation \cite{arandaEnhancingComputationalEfficiency2024}, and inverse pulse position modulation (i-PPM) and frequency shift keying (FSK) \cite{shenVehicularVisibleLight2018}.


\bigskip

\section{Experimental Methods}
\begin{figure}[t]
    \centering
    \includegraphics[width=\linewidth]{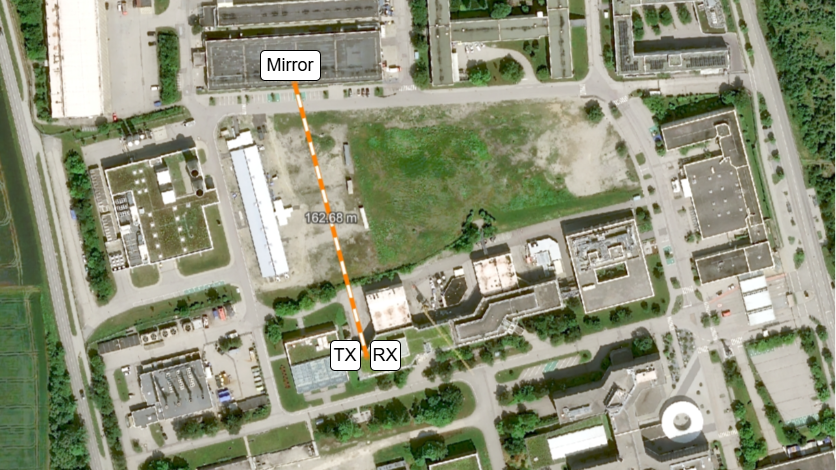}
    \caption{Satellite imagery showing the \qty{320}{\m} outdoor link path at an altitude of \qty{6.5}{\m}. Image rights: Airbus DS GmbH.}
    \label{fig:outdoor}
\end{figure}

Our setup employs a transmitter (TX) and receiver (RX) in a folded-path geometry across two buildings. A planar mirror extends the \qty{160}{\m} line-of-sight to an effective link range of \qty{320}{\m}, about \qty{6.5}{\m} over ground (\autoref{fig:outdoor}). 

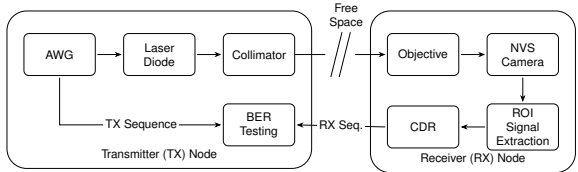
\begin{figure}[t]
    \centering
    \bigskip
    \resizebox{\linewidth}{!}{
    \begin{tikzpicture}[
    font=\large,
    block/.style={
        rectangle,
        draw,
        align=center, 
        rounded corners=5pt,
        minimum width=2.5cm,
        minimum height=1.5cm,
        text width=2.2cm, 
        inner sep=5pt
    },
    outerbox/.style={
        rectangle,
        draw,
        rounded corners=20pt,
        inner xsep=0.6cm,
        inner ysep=0.8cm
    },
    arrow/.style={
        -{Stealth[scale=1.0]},
        thick,
        shorten >=2pt,
        shorten <=2pt
    },
    freespaceline/.style={
        thick
    }
]

\node (AWG) [block] at (0, 0) {AWG};
\node (LD) [block, right=of AWG, node distance=1cm] {Laser\\Diode};
\node (Coll) [block, right=of LD, node distance=1cm] {Collimator};
\node (BERT) [block, below=of Coll, node distance=1.5cm] {BER\\Testing};

\node (Obj) [block] at (13, 0) {Objective};
\node (Cam) [block, right=of Obj, node distance=1cm] {NVS\\Camera};
\node (ROI) [block, below=of Cam, node distance=1.5cm] {ROI\\Signal\\Extraction};
\node (CDR) [block, left=of ROI, node distance=1cm] {CDR};

\node (tx_container) [outerbox, fit=(AWG) (LD) (Coll) (BERT)] {};
\node [anchor=south] at ([yshift=5pt]tx_container.south) {Transmitter (TX) Node};

\node (rx_container) [outerbox, fit=(Obj) (Cam) (ROI) (CDR)] {};
\node [anchor=south] at ([yshift=5pt]rx_container.south) {Receiver (RX) Node};


\draw [arrow] (AWG.east) -- (LD.west);
\draw [arrow] (LD.east) -- (Coll.west);
\draw [arrow] (AWG.south) |- node[pos=0.75, fill=white, inner sep=2pt] {TX Sequence} (BERT.west);

\draw [arrow] (Obj.east) -- (Cam.west);
\draw [arrow] (Cam.south) -- (ROI.north);
\draw [arrow] (ROI.west) -- (CDR.east);

\draw [arrow] (CDR.west) -- node[midway, fill=white, inner sep=2pt] {RX Seq.} (BERT.east);

\draw [thick] (Coll.east) -- (9.5, 0);
\draw [arrow] (10.5, 0) -- (Obj.west);

\draw [freespaceline] (10.1, 0.8) -- (9.7, -0.8);
\draw [freespaceline] (10.4, 0.8) -- (10.0, -0.8);
\node [anchor=south, align=center] at (10.25, 0.8) {Free\\Space};

\end{tikzpicture}
    }
    \caption{Schematic representation of the transmitter and receiver architecture with detail on transmission scheme.}
    \label{fig:exp}
\end{figure}

Data is transmitted via the optical beam using OOK-NRZ modulation.
To guarantee sufficient edge transitions for event generation 8b/10b encoding is employed, ensuring at least one rising or falling edge every five symbols \cite{widmerDCBalancedPartitionedBlock8B1OB}. 
At the TX node, an Arbitrary Wave Generator (AWG) generates a pseudo-random bit sequence (PRBS-9) \cite{nardoExperimentalCharacterizationOptical2026} to drive a near-infrared laser diode. The driving signal DC offset is set such that the minimum diode current is maintained above the lasing threshold, and the maximum optical powers remained within eye safety limits at all time.

The receiver containes a Neuromorphic Camera (Prophesee EVK4) equipped with the Sony IMX636 sensor with a $\qty{1280}{px} \times \qty{720}{px}$ resolution and \qty{4.86}{\micro\meter} pixel pitch. 
The sensor has a responsivity curve between \numrange{300}{1000}\unit{\nano\meter} and it is operated without any optical filters.
The camera is paired with a \qty{1.8}{\mm} focal length fish-eye objective, providing \qty{119.88}{\degree} horizontal- and \qty{88.37}{\degree} vertical-FoV. 
Real-time data processing is performed on an barebone computer; event stream batches of \qty{50}{\milli\s} are processed in under \qty{10}{\milli\s}. \autoref{fig:exp} shows a schematic representation of the transmitter and receiver architecture.

\begin{figure}[t]
    \centering
    \fcolorbox{black}{white}{
    \includegraphics[width=0.95\linewidth]{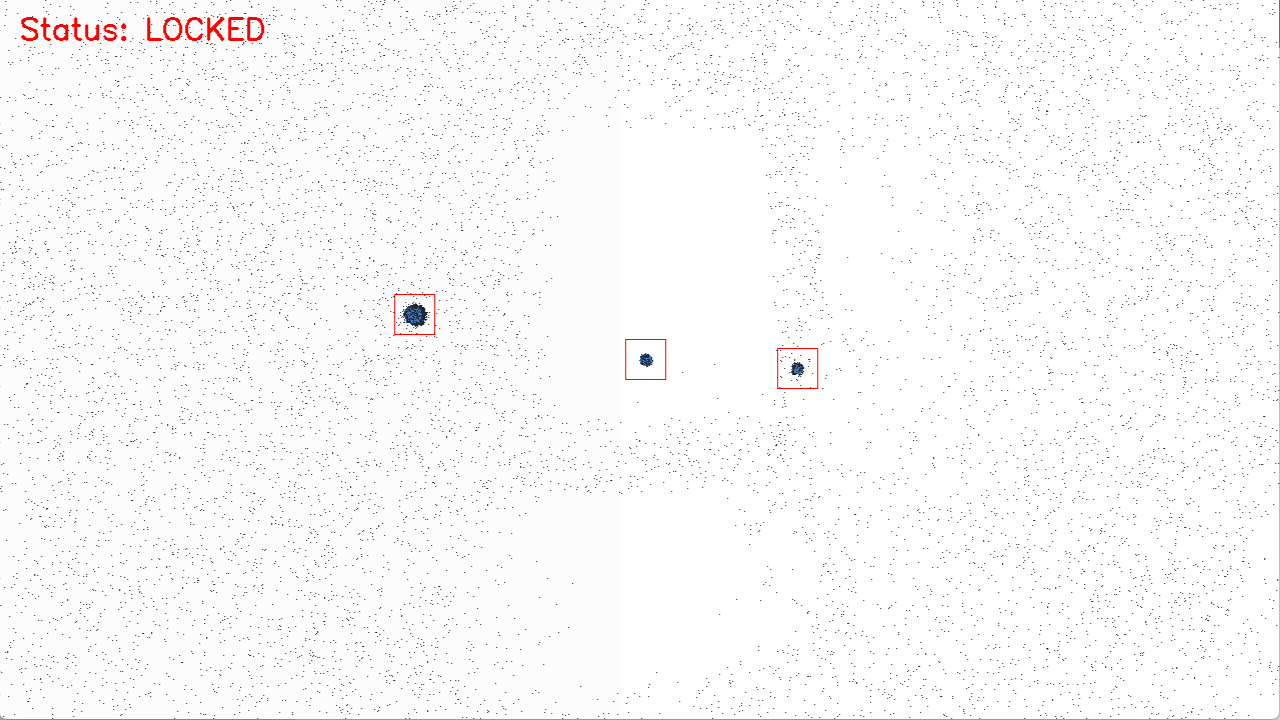}
    }
    \caption{Region of Interests (ROIs) of events generated by three \qty{80}{\kbps} modulated beams, \qty{240}{\kbps} combined. 
    }
    \label{fig:event}
\end{figure}

Our solution supports multiple simultaneous Regions of Interest (ROIs), \autoref{fig:event}, allowing for the parallel decoding of different beams. 
Unlike intensity-estimation methods (convert the event stream into grayscale images) \cite{scheerlinckContinuoustimeIntensityEstimation2018} found in previous work \cite{suMotionAwareOpticalCamera2025,suminoExperimentalDemonstrationEventbased2025}, our approach utilizes a binary quantization based on a majority decision within the corresponding ROI for the distinction of zeros and ones. When events are generated, the signal is extracted with temporal resolution of \qty{1}{\micro\s}.

From the resulting digital signal, strobe points are extracted at three samples-per-symbol. 
Clock and Data Recovery (CDR) is then performed using a digital Phase Locked Loop (PLL) based on the Bang-Bang phase detector \cite{alexanderClockRecoveryRandom1975} to determine an early/late clock and a Proportional-Integral (PI) controller \cite{razaviDesigningBangBangPLLs2009} to correct the phase error. The PI constants are chosen to reach convergence after 500 symbols. 
In this work, only the primary phase loop is implemented; consequently, the symbol rate must be known a priori. The implementation of a secondary loop for frequency acquisition is deferred to future work.

\section{Results}

First, the optimal symbol rate for the neuromorphic sensor was characterized in a controlled dark-room environment. 
In these conditions, the baseline background noise was measured at \qty{3000}{} events per second (\qty{}{evt/s}). 
To evaluate the sensor’s frequency response, a \qty{50}{\%} duty cycle square wave (simulating a repetitive $1010$ bit sequence) was used to modulate the laser diode source.

\begin{figure}[t]
    \centering
    \begin{tikzpicture}
\begin{axis}[
    width=7.5cm,
    height=4.5cm,
    tick label style={font=\tiny}, 
    label style={font=\tiny},
    xlabel={Modulation Bandwidth [\qty{}{\kilo\hertz}]},
    ylabel={Event-rate [\qty{}{kevt/s}]},
    xmin=0, xmax=350,
    ymin=0, ymax=2500,
    xtick={0, 50, 100, 150, 200, 250, 300, 350},
    ytick={0, 500, 1000, 1500, 2000, 2500},
    grid=both,
    grid style={line width=.1pt, draw=gray!10},
    major grid style={line width=.2pt, draw=gray!30},
    tick align=inside,
]

\addplot[
    color=black,
    mark=*,
    mark options={fill=black},
    thick
] coordinates {
    (2, 651)
    (4, 847)
    (8, 1195)
    (16, 1615)
    (32, 1853)
    (64, 2174)
    (100, 1691)
    (132, 970)
    (160, 420)
    (200, 137)
    (240, 34)
    (300, 9)
};

\end{axis}
\end{tikzpicture}
    \caption{Dark-room measured event rate (thousand events per second \qty{}{kevt/s}) versus modulation frequency for a \qty{50}{\%} duty cycle square wave, illustrating bandwidth limit.}
    \label{fig:bandwidth_saturation}
\end{figure}
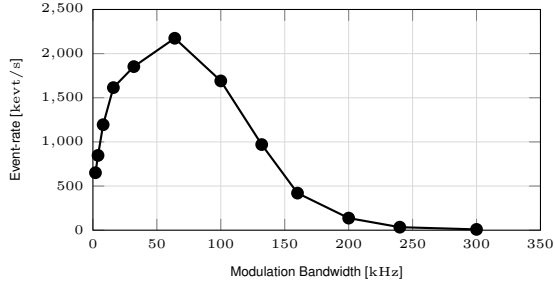

As shown in \autoref{fig:bandwidth_saturation}, the effective event throughput does not reach the theoretical \qty{1}{\mega\hertz} limit of the IMX636 sensor. 
Despite fine-tuning the camera biases to minimize the refractory period of each pixel, a measurable dead-time persists, limiting the peak event generation rate. Furthermore, the sensor's readout architecture introduces a bottleneck: while pixels in the same row receive identical timestamps, the interface electronics can only process \num{~100} lines per \qty{}{\micro\s}. Consequently, a full-frame readout (\num{720} vertical lines) caused by multiple active beams requires \qty{8}{\micro\s}, i.e. a maximum theoretical bandwidth of \qty{125}{\kilo\hertz} close to our short range maximum achieved data rate (\qty{106}{\kbps}). 

Real-time outdoor experiments were conducted under daytime solar illumination in a static scenario. Background noise events across the sensor array further consume available bandwidth, necessitating bias adjustment to equalize the sensitivity for rising and falling edges (positive and negative polarities).
The system achieved a data rate of \qty{106}{\kilo\bit\per\second} ($BER$ of \num{8.28E-04
}) at a distance of \qty{1}{\m} (indoor) and \qty{80}{\kilo\bit\per\second} ($BER$ of \num{6.55E-04}) at a distance of \qty{320}{\m} (outdoor). 
The \qty{120}{\degree} FoV objective enabled the concurrent operation of multiple spatial communication channels with only a marginal degradation, 
of the performance of individual links.
As shown in \autoref{fig:event}, three emitters, from three different TX nodes, are transmitting at \qty{80}{\kbps} each, for an aggregate \qty{240}{\kbps}.

\begin{figure}[t]
    \centering
    \includegraphics[width=1\linewidth]{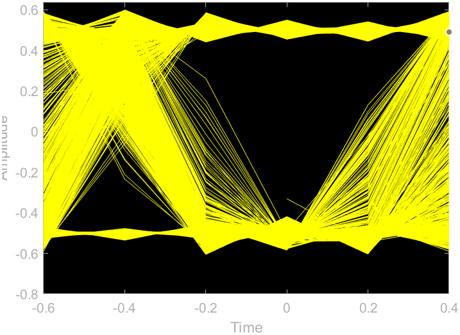}
    \caption{Recovered digital eye diagram of the RX signal normalized between \num{-0.5} and \num{0.5} ($BER = \num{6.55E-04}$).}
    \label{fig:eye}
\end{figure}

Regarding CDR, the recovered digital eye diagram following the PLL stage is presented in \autoref{fig:eye}. 
For visualization, the signal is resampled to \num{6} samples-per-symbol and normalized between \num{-0.5} and \num{0.5}. 
As a final remark, the $BER$ at extended distances can be improved by increasing the TX laser diode’s optical power or choosing a telescopic objective as in \cite{suminoEventIntervalModulation2025}. While operating at a constant eye-safe power level, for increasing distances, we had to decrease the symbol rate to maintain a fixed $BER$, mainly due to the effect of scintillation from atmospheric turbulence.

\section{Conclusion}
This work demonstrated a real-time event-based Optical Camera Communication (event-OCC) system utilizing a wide \qty{120}{\degree} Field-of-View (FoV) neuromorphic sensor. The experimental results characterized an outdoor link over a distance of \qty{320}{\m} in daylight conditions, without cloud cover and without the requirement for optical bandpass filters. We employed an OOK-NRZ modulation with 8b/10b encoding and a digital Phase-Locked Loop based on a digital Bang-Bang phase detector for the clock recovery. 
In a single node operation, the system achieved a data rate of \qty{106}{\kbps} at \qty{1}{\m} (indoor) and \qty{80}{\kbps} at \qty{320}{\m} (outdoor), while maintaining a Bit Error Ratio ($BER$) below $10^{-3}$. 
In multi-node operation, using three parallel incoming beams consisting of the \qty{320}{\m} outdoor link and several \qty{1}{\m} indoor links, the system achieved a accumulated data rate of \qty{240}{\kbps} at an average $BER$ of \num{1.69E-03}. These results establish a new baseline for both event-OCC and Free Space Optical Communication in particular in application requiring moderate data rates and large FoV.

While the communication bandwidth of a single beam is limited by the temporal resolution and the readout architecture of the employed neuromorphic sensors (Sony IMX636) to an estimated \qty{125}{\kilo\hertz} upper bound, further throughput increases can be achieved through parallelization. The \qty{120}{\degree} FoV provided by the optical system inherently supports Spatial Division Multiplexing (SDM). By defining multiple independent Regions of Interest across the sensor array, the aggregate data rate can be scaled linearly with the number of transmitted beams without increasing the complexity of the receiver hardware.
In addition to SDM, future work could investigate Wavelength Division Multiplexing (WDM) through Diffractive Optical Element at the receiver aperture to spatially separate incoming signals based on their wavelength.

Such advancements, combined with the wide FoV and high dynamic range of neuromorphic sensors, position event-OCC as a promising technology for low-cost, low-size, low-weight, low-power and reliable optical wireless communication in mobile, vehicular and aerospace applications.

\clearpage

\defbibnote{myprenote}{%
Citations must be easy and quick to find. More precisely:
\begin{itemize}
    \item Please list all the authors. 
    \item The title must be given in full length. 
    \item Journal and conference names should not be abbreviated but rather given in full length.
    \item The DOI number should be added incl. a link.
\end{itemize}
}
\printbibliography

\vspace{-4mm}

\end{document}